# Observation of the chiral anomaly induced negative magneto-resistance in 3D Weyl semi-metal TaAs

Xiaochun Huang[1,§], Lingxiao Zhao[1,§], Yujia Long[1], Peipei Wang[1], Dong Chen[1], Zhanhai Yang[1], Hui Liang[1], Mianqi Xue[1], Hongming Weng[1,2], Zhong Fang[1,2], Xi Dai[1,2] and Genfu Chen[1,2,*]

[1]*Institute of Physics and Beijing National Laboratory for Condensed Matter Physics, Chinese Academy of Sciences, Beijing 100190, China*

[2]*Collaborative Innovation Center of Quantum Matter, Beijing, 100190, China*

§*: These authors contributed equally to this work*

\**: Correspondence should be addressed to: gfchen@iphy.ac.cn*

## Abstract

Weyl semi-metal is the three dimensional analog of graphene. According to the quantum field theory, the appearance of Weyl points near the Fermi level will cause novel transport phenomena related to chiral anomaly. In the present paper, we report the first experimental evidence for the long-anticipated negative magneto-resistance generated by the chiral anomaly in a newly predicted time-reversal invariant Weyl semi-metal material TaAs. Clear Shubnikov de Haas oscillations (SdH) have been detected starting from very weak magnetic field. Analysis of the SdH peaks gives the Berry phase accumulated along the cyclotron orbits to be $\pi$, indicating the existence of Weyl points.

When two non-degenerate bands cross in three dimensional momentum space, the crossing points are called Weyl points, which can be viewed as magnetic monopoles (*1*) or topological defects (*2*) in band structure, like "knots" on a rope. Near Weyl points, the low energy physics can be described by Weyl equations (*3*) with distinct chirality (either left- or right-handed), which mimics the relativistic field theory in particle physics. On lattice system, Weyl points always appear in pairs with opposite chirality and are topologically stable against perturbations that keep translational symmetry (*4-7*). If two weyl points with opposite chirality meet in the momentum space, they will generally annihilate each other, but may also be stabilized as 3D Dirac points by additional (such as crystalline) symmetry (*8-11*). For materials with Weyl points located near the Fermi level, called as Weyl semi-metals (WSMs), exotic low energy physics will be expected, such as the Fermi arcs on the surfaces (*5,6*), and the chiral-anomaly induced quantum transport (*12-15*). Recently, 3D Dirac semimetals, $Na_3Bi$ and $Cd_3As_2$, have been theoretically predicted (*9,10*) and experimentally confirmed (*16-20*), while WSM are still waiting for its experimental verification in spite of various theoretical proposals (*5,6,21-25*).

The anomalous DC transport properties are important consequence of the topological band structure (*14,26,27*). In topological insulators (TI), the transport properties are dominated by the topological surfaces states (SS), where the lack of back scattering caused by the unique spin structure of the SS leads to the weak anti-localization (WAL) behavior. While in Weyl semi-metals, the bulk states are semi-metallic and dominate the DC transport. In relativistic field theory, for a continue system described by Weyl equation, chiral anomaly can be understood as the non-conservation of the particle number with given chirality, which only happens under the presence of parallel magnetic and electric fields (*12*). For any realistic lattice system, the chiral anomaly then manifests itself in the inter-valley pumping of the electrons between Weyl points with opposite chirality. In the non-interacting case, the chiral anomaly can be simply ascribed to the zeroth Landau levels, which are chiral and have opposite sign of the velocity for states around Weyl points with opposite chirality, as shown in Fig. 1D (*12*). The additional electric field parallel to the magnetic field will then generate imbalance between two chiral modes leading to electric current which can only be balanced by inter-valley scattering. Considering the facts that for clean samples the inter-valley scattering time is extremely long and the degeneracy of the Landau level is proportional to the magnetic field strength, the chiral anomaly in WSM will in general lead to negative magneto-resistance $(MR = {(\rho(H) - \rho(0))}\big/{\rho(0)})$ when the magnetic field is parallel to the current. On the other hand, for ordinary metal or semiconductors the MR is weak, positive and usually not very sensitive to the magnetic field direction. Therefore, the negative and highly anisotropic MR has been regarded as the most prominent signatures in transport for the chiral anomaly and indicates the existence of 3D Weyl points. Besides, chiral anomaly can also generate other fascinating phenomena, i.e. the anomalous Hall effect and nonlocal transport properties (*6,13*).

Using first principle calculations, Weng *et al.* (*24*) predicted that a family of binary compounds represented by TaAs are time-reversal invariant 3D WSMs with a dozen pairs of Weyl nodes which are generated by the absence of inversion center. Materials in TaAs family are completely stoichiometric and nonmagnetic, providing an almost ideal platform for the study of chiral anomaly in WSM. In this work, we perform transport studies of the TaAs single crystal down to 1.8 K with magnetic field up to 9

T. Ultrahigh mobility ($\mu_e \approx 1.8 \times 10^5$ cm$^2$V$^{-1}$s$^{-1}$ at 10 K) has been found with multiband character. Extremely large positive MR ($\approx$ 80000% at 1.8 K in a field of 9 T) has been discovered for magnetic field perpendicular to the current (or the external electric field). When the magnetic field is rotated to be parallel to the current, notable negative MR has been observed, demonstrating the chiral anomaly effects in this particular material. Strong SdH oscillations have also been found from very low magnetic field, from which two sets of oscillation frequencies can be extracted indicating two types of carriers, in good consistence with first-principles calculations.

TaAs crystallizes in a body-centered-tetragonal NbAs type structure with nonsymmorphic space group of $I4_1md$, in which $c$-axis is perpendicular to the $ab$-plane (See Fig.1A). The lattice parameters are $a=b$=3.4348 Å and $c$=11.641 Å (*28*). Due to the lack of inversion symmetry, first principle calculations predict a dozen pairs of Wely points in the Brillouin zone (BZ) (*24*). A schematic diagram of theoretical predicted Weyl nodes projected on the (001) facet can be seen in Fig. 1B. Single crystals of TaAs studied in this work are synthesized via chemical vapor transport method (See Method). Figure 1C shows the X-ray diffraction from a TaAs single crystal oriented with the scattering vector perpendicular to the (001) plane. The inset is the morphology of a representative sample looking down the [001] direction. Samples for electric transport measurement are polished to a flat thin strip, and the largest facet is the (001) plane. The electric current is always applied parallel to the (001) plane along the $a$ or $b$ axis in our studies. Magneto-resistance and Hall resistivity measurements with Four-Point Probe and Alternating Current Transport methods are carried out in a Quantum Design Physical Property Measurement System (PPMS).

Figure 2 presents the MR measured at 1.8 K by tilting the magnetic field ($B$) at an angle ($\theta$) with respect to the electric current ($I$). As shown in Fig. 2A, when the magnetic field is applied perpendicular to the current ($B \perp I$, $\theta$=0 °), a positive MR of up to 80000％ is observed. At low magnetic fields, MR exhibits quadratic field dependence and soon changed to almost linear dependence at very low field without any trend towards saturation up to 9 T. This giant conventional transverse MR strongly relies on $\theta$ and decreases considerably with increasing $\theta$. When the magnetic field is rotated to be parallel to the electric current ($\theta$=90 °), we get a negative MR, a strong evidence of Wely fermions in TaAs. Elaborate measurements at different angles around $\theta$=90 °are implemented and presented in Fig. 2B. As shown in the main panel, rotating $\theta$ from 87 °to 91.8 °, negative MR is arising in the cases of $\theta$ between 88 °and 91 °, and most obviously at 90 °($B$//$I$) with a value of -30％. This can also be intuitively viewed as a consequence of the steep downturn of MR in the magnetic field range from 1 T to 6 T (and -1 T to -6 T). In this range, for clarity, the minima of MR curves at different angles are listed in the inset of Fig. 2B. The largest value as expected occurs at $\theta$=90 °.

The origin of the negative MR in TaAs can be explained by the chiral anomaly under the semi-classical approximation (*29*). Under the magnetic field, the low energy states near the Weyl points will reorganize to form Landau states for the motion perpendicular to the field and leave the momentum $\boldsymbol{k}$ parallel to the field still a good quantum number. As shown in Fig. 1D, the zeroth landau levels are chiral with the chirality determined by that of the Weyl point. With the additional electric field along the same direction, the equation of motion for the electrons on the chiral modes gives $\hbar \frac{dk}{dt} = -e\boldsymbol{E}$, which adiabatically pumps electrons from one valley to another one with opposite chirality. Solving the corresponding Boltzmann equation under the semi-classical approximation gives chiral anomaly contributed conductivity as

$$\sigma_a = \frac{e^3 v_f^3}{4\pi^2 \hbar \mu^2 c} B^2,$$

where $\tau$ is the inter valley scattering time, $v_f$ is the Fermi velocity near the Weyl points and $\mu$ denotes the chemical potential measured from the energy of the Weyl points. The above chiral part of the conductivity increasing quadraticly with magnetic field $B$ leads to negative MR, which has the maximum effect with *E* parallel to *B*. Of course the total conductivity of the system will also include other contributions from the non-chiral states as well, which may weaken the negative MR effect or even overwhelm it if the non-chiral part dominates the DC transport. Therefore in order to see the chiral negative MR, the high quality sample with chemical potential close enough to the Weyl point is crucial. In this work, this can be roughly recognized by the coexistence of SdH oscillations and giant transverse MR (See Fig. 2A), which usually implied small Fermi surfaces around Fermi level (*30*). Further discussions in the following will give quantitative analyses. Measurements are also implemented by tilting the magnetic field with respect to the (001) facet but keeping $B \perp I$ (See Fig. S1b). As expected, no evidence of negative MR phenomenon has been detected and adds to growing evidence that the negative MR origins from the chiral term *E* · *B*.

Near zero field, all the data in Fig. 2B show sharp dips, which may be attributed to the WAL effect stemming from the strong spin-orbit interactions (*27*), which dominates the transport behavior of the non-chiral states. For magnetic field higher than 9 T, the MR will change sign to be positive again. This behavior is very similar to the situation in Bi$_x$Sb$_{(1-x)}$ (*27*). Since the separation between the Weyl points in *k*-space is about 3-8% of the zone boundary (*24*), which is quite big comparing with the energy scale of the Zeeman coupling, it is very unlikely that the Weyl points will annihilate in pairs in high magnetic field. One possible explanation is due to the Coulomb interaction among the electrons occupying the chiral states. Since the degeneracy of the chiral states as well as the density of states at the Fermi level goes linearly with the magnetic field, eventually the system will approach a spin-density-wave (SDW) like instability under Coulomb interaction (*31*). Then at finite temperature, the strong SDW fluctuation provides another scattering channel which can be greatly enhanced in high field and may give the positive MR in the high field region.

Figure 2C shows the magnetic field dependence of original resistivity data at different temperatures for $\theta$=90 °. The negative MR is suppressed with increasing temperature, and ultimately disappeared at higher temperature. Variations of MR as a function of the perpendicular component of the magnetic field, $B\cos\theta$, are studied and shown in Fig. 2D. The misalignment of the curves indicates a 3D nature of the electronic structure in TaAs. This is reconfirmed by the explicit SdH oscillations in the whole angle range (See Fig. 2A).

Figure 3A shows the magnetic field dependence of Hall resistivity $\rho_{xy}(B_z)$ measured at various temperatures. At low temperature, the negative slope in high magnetic fields indicates that the electrons dominate the main transport processes. However, in low fields the curve tends to be flat. Remarkably, we note that, the negative slope of Hall resistivity changes to positive at higher temperature, implying the carriers dominating the conduction mechanism transformed to hole-type. The transition temperature as shown in Fig. 3B is about 100 K. At this temperature, not only the slope, but also the values of Hall resistivity change signs. This signified the possibility of the coexistence of high-mobility electrons with low-mobility holes (*30,32,33*). Using two-carrier model (*32,33*), both longitudinal conductivity $\sigma_{xx}$ and Hall conductivity $\sigma_{xy}$ are fitted to estimate the mobility and the concentration of electrons ($\mu_e$, $n_e$) and holes ($\mu_h$, $n_h$) at various temperatures. The results of the temperature dependent fitting parameters are summarized in Fig. 3C and Table S1-2. In the low temperature regime, the concentration of two carriers is comparable, while the mobility of electrons is one order of magnitude larger than that of holes. For the electrons, an extremely high value of $\mu_e{\approx}1.8{\times}10^5$ cm$^2$V$^{-1}$s$^{-1}$ is obtained at T=10 K. Above 80 K, however, $\mu_e$ starts to drop dramatically by almost two order of the magnitude, but $\mu_h$ shows a weak temperature dependence, suggesting that the holes dominate the transport properties at higher temperature, which is in agreement with the results derived from the Hall resistivity. The inset of Fig. 3C shows the temperature dependence of $\mu_e$ and $\mu_h$ obtained from the longitudinal conductivity $\sigma_{xx}$ fitting procedures, in the low temperature regime, which is consistent with that of $\mu_e$, and $\mu_h$ deduced from $\sigma_{xy}$ fitting procedures. The two-carrier model and fitting details are given in the Supplementary Material.

Magnetic field dependence of resistivity $\rho_{xx}$ at various temperatures is measured at $\theta$=0 ° and shown in Fig. 3D. Obvious SdH oscillations at lower temperatures are suppressed with increasing temperature and ultimately eliminated around 20 K. Fourier transformation of the oscillations at 1.8 K shows two major frequencies at 6 and 16 T. The cross-sectional area of the Fermi surface (FS) $A_F$ can be obtained according to the Onsager relation $F = \left( {}^{\emptyset_0}\!\big/\!{}_{2\pi^2} \right) A_F$, where $F$ is oscillation frequency and $\emptyset_0$ is flux quantum. The calculated values of $5.721{\times}10^{-4}$ Å$^{-2}$ and $15.26{\times}10^{-4}$ Å$^{-2}$ are only $1.713 \times 10^{-6}\%$ and $4.568 \times 10^{-6}\%$ of the cross-sectional area of the first BZ, respectively. Figure 3E shows the plots of the oscillatory components $\Delta\rho_{xx}$ versus inverse field $1/B$ at 1.8 K. The peaks marked by Landau index $n$ are mainly stemming from the oscillations with the frequency of 16 T, and a magnetic field of 7 T drives the sample to the $2^{nd}$ Landau level. According to the well-known semi-classical Lifshitz-Onsager relation: $A_F(\frac{h}{2\pi eB}) = 2\pi(n + \gamma)$, the phase offset $\gamma$ of the quantum oscillations should to be 0 (or 1) for 2D Dirac fermion and -1/8 (or 7/8) for 3D case with additional $\pi$ Berry phase (*34-36*). As shown in the inset of Fig. 3E, a linear fitting with constraint frequency of 16 T yields an intercept of -0.08, which is very close to the value of -1/8, signaling the 3D Dirac fermion behaviors, and implying the FS associated with 16 T quantum oscillations is enclosing a Weyl point. The other oscillations associated with 6 T are too weak in amplitude to be labeled. We note also that $\rho_{xx}(B)$ has transition from quadratic field dependence at low fields to linear dependence at higher fields. The crossover field strength increases monotonically with the temperature as introduced in more detail in the supplementary (See Fig. S4). In Fig. 3F the temperature dependence of resistivity $\rho_{xx}$ at $\theta=0\,^{\circ}$ is plotted. In zero magnetic field, TaAs exhibits a metallic behavior down to 1.8 K (See the inset of Fig. 3F). The applied magnetic field not only significantly increases the resistivity, but also stimulates a crossover from metallic to insulator like behavior, which may be related to the formation of the Landau levels under magnetic field.

The FS cross sections determined by the quantum oscillation is roughly consistent with the first principle calculations showing that there are three kinds of Fermi surfaces in the first BZ (Fig. 4A and 4B). The clearly visible two are banana-like hole-pockets and dates-like electron-pockets. The eight banana-like hole-pockets are from band 2 as shown in Fig. 4C. The 16 dates-like electron pockets are enclosing the Weyl nodes in the $k_z$=0.592$\pi$ plane (W1) (*24*). These nodal points are around 21 meV below the Fermi level. The band structures around W1 along $k_x$, $k_y$ and $k_z$ are shown in Fig. 4D-F, indicating W1 are not strongly anisotropic with averaged Fermi velocity along three directions are 2.025, 2.387 and 2.231 eV•Å, respectively. The Weyl node in $k_z$=0 plane, W2, is around 2 meV higher than the Fermi level, which contributes to the third type Fermi surface, which is too tiny and can be can marked by dots in Fig. 4B. W2 is strongly anisotropic as shown by its band structures along $k_x$, $k_y$ and $k_z$ in Fig4. G-I. The Fermi velocities along them are 1.669, 2.835 and 0.273 eV•Å, respectively. Combining the information obtained from both quantum oscillation experiments and first principle calculations, we can conclude that it is likely the FS associated with 16 T oscillating frequency encloses the Weyl points W1 and dominate the DC transport in TaAs, which on the other hand strengthens our conclusion that the negative MR observed in this material is truly due to chiral anomaly.

In summary, the transport properties of proposed WSM TaAs have been studied in detail. Our results suggest that there are both *n* and *p* types of carrier in TaAs, which nearly compensate to each other. Further analyses of the Hall effect data through two-carrier fittings obtained the electronic motilities at different temperatures and gives out an unexpected high value of $\mu_e \approx 1.8 \times 10^5$ cm$^2$V$^{-1}$s$^{-1}$ for the electrons at 10 K. Giant linear MR as high as 80000% at 9T has been found with magnetic field perpendicular to the current, which is quite similar with the behavior in Dirac semi-metal Cd$_3$As$_2$. When the external magnetic field is rotated to be parallel with the current, the MR becomes negative between 1-9 T, with the most negative value to be -30%. This unusual negative MR is the first evidence for the chiral anomaly associated with the Weyl points in TaAs. The π Berry phase acquired by electrons in cyclotron orbits is another strong evidence of Weyl points. Thus, TaAs is the first nonmagnetic compound confirmed by the experiments to host WSM state, which has paved the way for more experimental studies in this completely new research territory.

During preparation of this manuscript the authors became aware of a related work by Zhang *et al.* posted on arXiv (arXiv:1502.00251(2015)).

## Method

Single crystals of TaAs were grown by chemical vapor transport. A previously reacted polycrystalline TaAs was filled in the quartz ampoule using 2 mg/cm$^3$ of Iodine as the transporting agent. After evacuating and sealing, the ampoule was kept at the growth temperature for three weeks. Large polyhedral crystals with dimensions up to 1.5 mm are obtained in a temperature field of $\Delta T$ = 1150 ℃-1000 ℃. The as grown crystals were characterized by X-ray diffraction (XRD) using PANalytical diffractometer with Cu Kα radiation at room temperature. The single-crystal X-ray diffraction determined crystal growth orientation. The average stoichiometry was determined by energy-dispersive X-ray (EDX) spectroscopy. No I$_2$ doping was detected. For magneto-resistance and Hall measurements, a platelike specimen (1×0.3×0.08 mm$^3$) was prepared from the as grown crystal by polishing. The magneto-resistance was measured with the four-point probe method in a Quantum Design PPMS, and the Hall coefficient measurement was done using a five-probe technique. For MR (or Hall resistivity) measurements, any Hall (or resistive) voltages due to misalignment of the voltage leads could be corrected by reversing the direction of the magnetic field.

First principle calculations have been performed by using OpenMX (*37*) software package. The choice of pseudo atomic orbital basis set with Ta9.0-s2p2d2f1 and As9.0-s2p2d1, the pseudo-potential and the sampling of BZ ($10 \times 10 \times 10$ $k$-grid) have been checked. The exchange-correlation functional within generalized gradient approximation parameterized by Perdew, Burke, and Ernzerhof has been used (*38*). The optimized lattice constants $a=b=$3.4824 Å, $c=$11.8038 Å and atomic sites are in agreement with the experimental values.

**Acknowledgements**

The authors wish to thank L. Lu, Z. P. Hou and C. Ren for their fruitful discussions and helpful comments. This work was supported by National Basic Research Program of China 973 Program (Grant No. 2015CB921303, 2011CBA00108 and No. 2013CB921700), the "Strategic Priority Research Program (B)" of the Chinese Academy of Sciences (Grant No. XDB01020300 and No. XDB07020100) and the National Natural Science Foundation of China.

**Author Contributions**

G. C., Z. F. and X. D. proposed and designed the project, G. C. and L. Z. prepared the samples, G. C., X. H. and L. Z. carried out the resistivity, magnetic susceptibility, Energy dispersive X-ray and X-ray diffraction measurements. Y. L., P. W., D. C., Z. Y., H. L. and M. X. provided assistance with sample synthesis. H. W., Z. F. and X. D. did theoretical calculations. All the authors analyzed the data and discussed the results. G. C., X. H., H. W. Z. F. and X. D. wrote the paper.

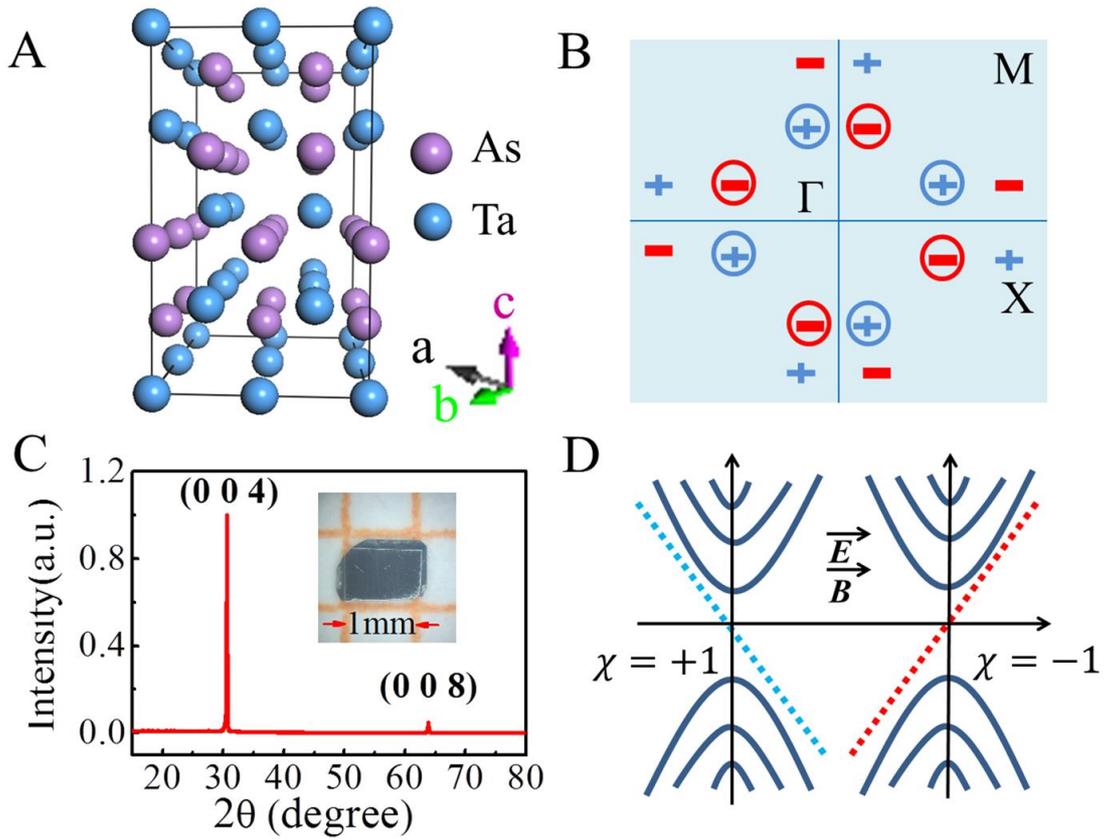

**Figure 1: Structure and symmetry of TaAs single crystal.** (A) The crystal structure of TaAs with nonsymmorphic space group of $I4_1md$. Blue and violet balls represent Ta atom and As atom, respectively. (**B**) Schematic diagram of a dozen pairs of Weyl points projected on the (001) facet. "+" and "-" denote Weyl points with positive and negative chiralities, respectively. The circles represent there are two Weyl points with same chirality projected on the same point in the (001) facet. Γ, X and M are the high symmetry points in the Brillouin zone. (**C**) X-ray diffraction pattern of TaAs single crystal. Inset shows optical image of a typical sample at the millimeter scale. (**D**) Schematic diagram of bulk Landau levels of a pair of Weyl nodes. The dotted lines represent the zeroth quantum Landau Level with "+" (blue) and "-" (red) chiralities in a magnetic field parallel to electric current.

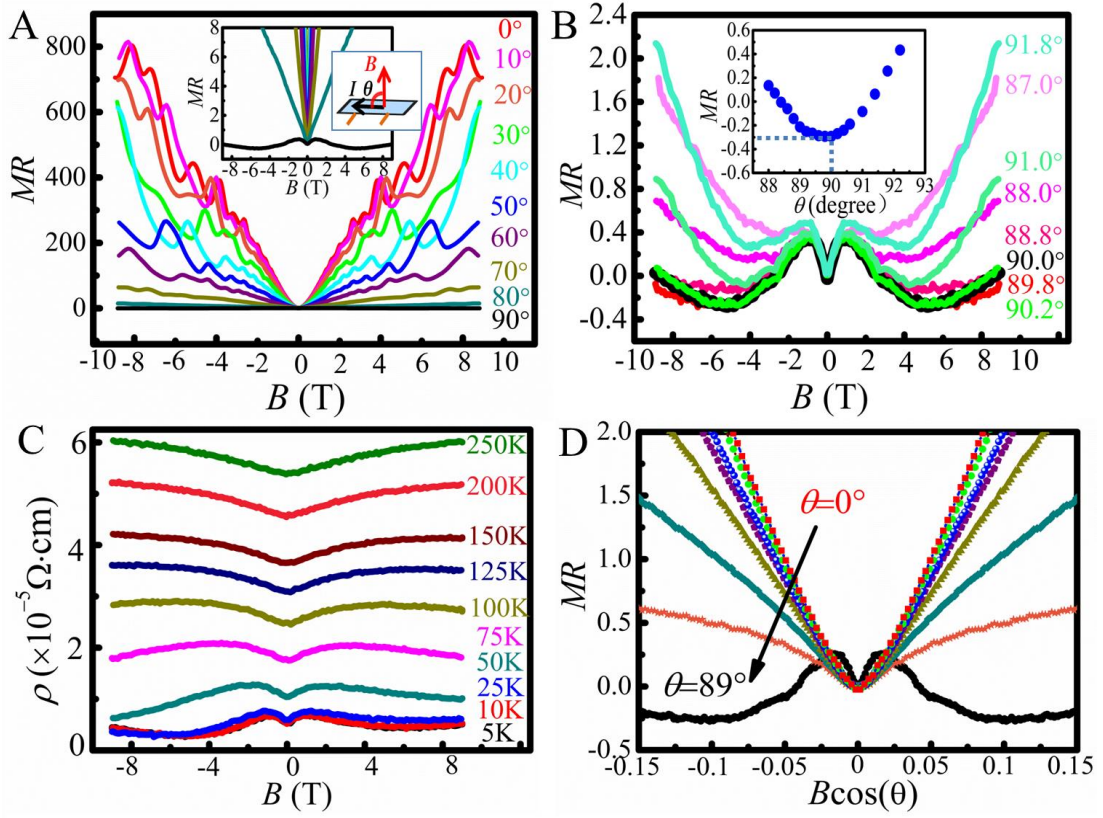

**Figur 2: Angular and field dependence of MR in TaAs single crystal at 1.8 K.** (**A**) Magneto-resistance with magnetic field (*B*) from perpendicular (*θ*=0 °) to parallel (*θ*=90 °) to the electric current (*I*). The inset zooms in on the lower MR part, showing negative MR at *θ*=90 ° (longitudinal negative MR), and depicts the correspondingly measurement configurations. (**B**) Magneto-resistance measured in different rotating angles around *θ*=90 ° with the interval of every 0.2 °. The negative MR appeared at a narrow region around *θ*=90 °, and most obviously when *B//I*. Either positive or negative deviations from 90 ° would degenerate and ultimately kill the negative MR in the whole range of magnetic field. Inset: the minima of MR curves at different angles (88 °-92.2 °) in a magnetic field from 1 to 6 T. (**C**) Magnetic field dependence of the original resistivity data at different temperatures for *θ*=90 °. (**D**) Magneto-resistance in the perpendicular magnetic field component of field, *B*cos*θ*. The misalignment indicates the 3D nature of the electronic states.

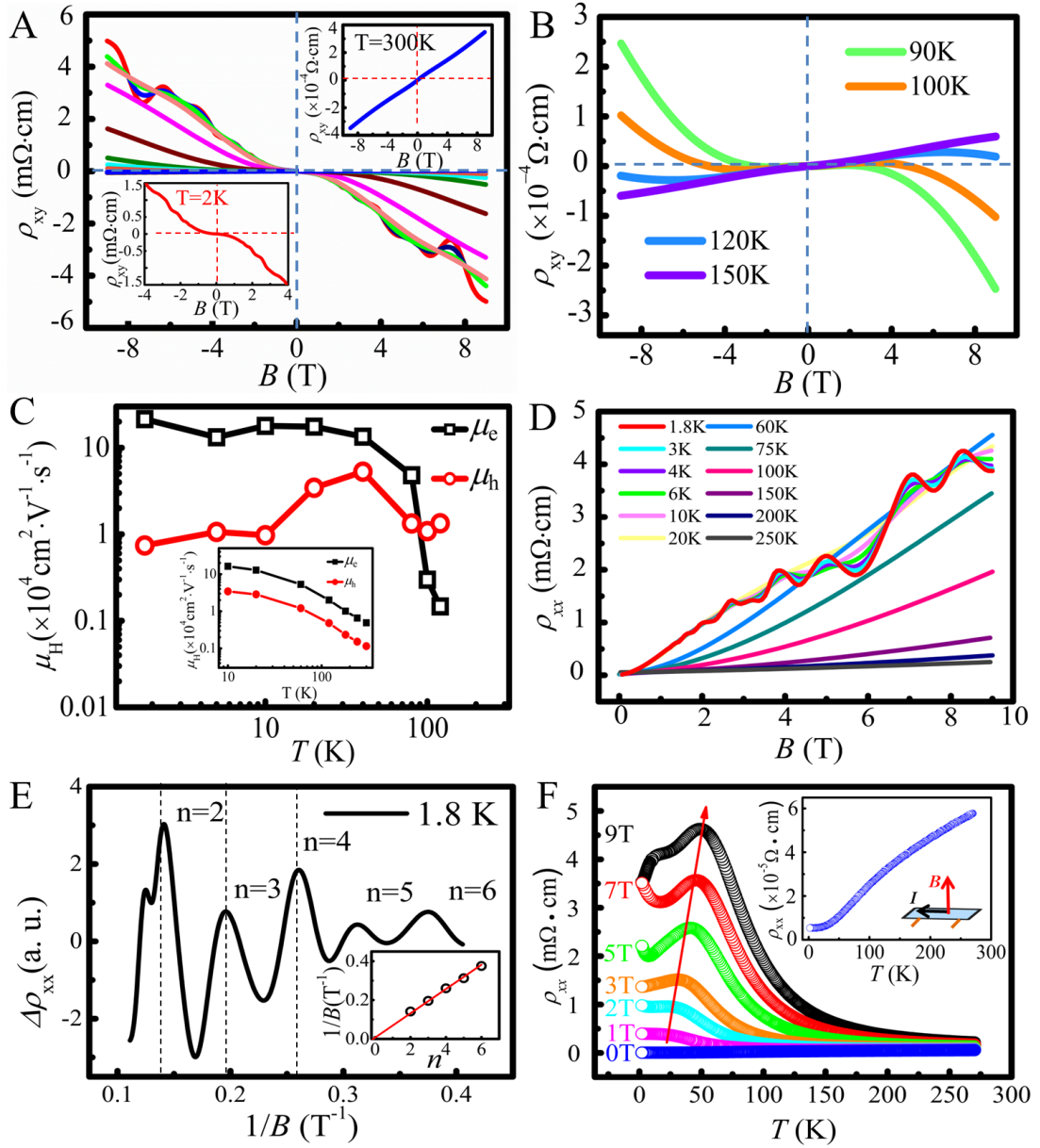

**Figure 3: Temperature dependence of Hall resistivity and resistivity for TaAs.** (**A**) Hall resistivity measured at various temperatures from 2 to 300 K. The upper right inset and the lower left inset show the Hall resistivity at 300 and 2 K, respectively. The obvious SdH oscillation demonstrates the high quality of the sample. (**B**) Hall resistivity at T = 90, 100, 120 and 150 K. At T = 100 K, both the Hall resistivity and its slope change signs, signaling the coexistence of two types of carriers in TaAs. (**C**) Temperature dependence of carrier-mobility $\mu_e$ and $\mu_h$ for electrons and holes deduced

by two-carrier model. Main panel: fitting with $\sigma_{xy}$; Inset: fitting with $\sigma_{xx}$. (**D**) Magnetic field dependence of resistivity with $\theta$=0 ° at representative temperatures. (**E**) The high-field oscillatory component plotted against inverse field $1/B$ at 1.8 K. Inset: Landau index $n$ with respect to $1/B$. The circles are the experimental data and the red line expresses the theoretical fitting. (**F**) The temperature dependence of resistivity in magnetic field perpendicular to the electric current. The red arrow indicates that the resistivity peak moves to high temperature under much higher magnetic fields. The inset gives the measurement configuration, and zooms in on the case of 0 T.

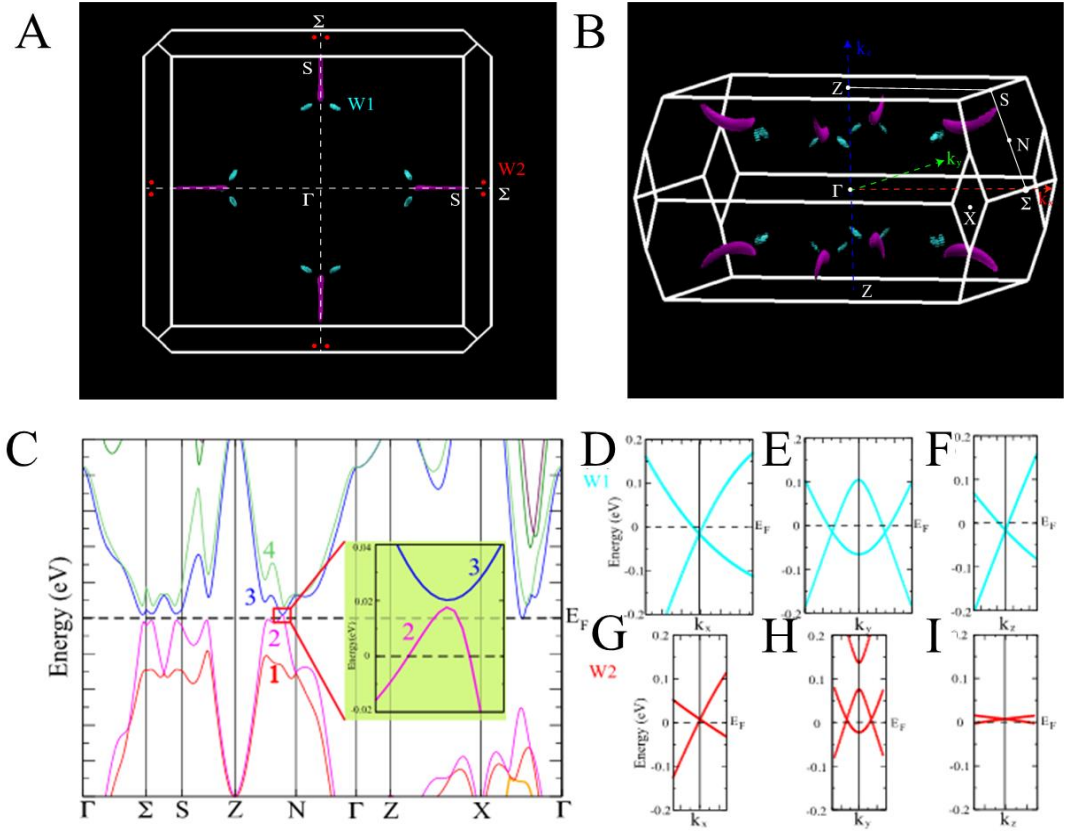

**Figure 4: Theoretical analyzing of Fermi surfaces in TaAs.** (**A and B**) First-principles calculated Fermi surfaces in top view and side view, respectively. (**C**) The band structure with SOC included. (**D**)-(**F**) and (**G**)-(**I**) Band structure around Weyl nodes in $k_z=0.592\pi$ (W1) and $k_z=0$ (W2) plane along $k_x$, $k_y$, and $k_z$ direction, respectively.